\documentclass{iopart}

\usepackage{epsfig}

\begin{document}
\title[Background of GWs from pre-galactic black hole
formation]{Background of gravitational waves from pre-galactic
black hole formation}
\author[J C N de Araujo \etal]{J C N de Araujo, O D Miranda and O D Aguiar}
\address{Instituto Nacional de Pesquisas Espaciais - Divis\~ao de
Astrof\'\i sica \\ Av. dos Astronautas 1758, S\~ao Jos\'e dos
Campos, 12227-010 SP, Brazil}
%
%
\begin{abstract}
We study the generation of a gravitational wave (GW) background
produced from a population of core-collapse supernovae, which form
black holes in scenarios of structure formation of the Universe.
We obtain, for example, that a pre-galactic population of black
holes, formed at redshifts $z\simeq 30-10$, could generate a
stochastic GW background with a maximum amplitude of $h_{\rm
BG}\simeq 10^{-24}$ in the frequency band $\nu_{\rm obs}\simeq
30-470 {\rm Hz}$ (considering a maximum efficiency of generation
of GWs, namely, $\varepsilon_{\rm GW}=7\times 10^{-4}$). In
particular, we discuss what astrophysical information could be
obtained from a positive, or even a negative, detection of such a
GW background produced in scenarios  such as those studied here.
One of them is the possibility of obtaining the initial and final
redshifts of the emission period from the observed spectrum of
GWs.
\end{abstract}
%


\pacs{0430D, 9760L}


\section{Introduction}

Because of the fact that gravitational waves (GWs) are produced by
a large variety of astrophysical sources and cosmological
phenomena, it is quite probable that the Universe is pervaded by a
background of such waves. A variety of binary stars (ordinary,
compact or combinations of them), Population III stars, phase
transitions in the early Universe and cosmic strings are examples
of sources that could produce such a putative background of GWs
(Thorne 1987, Blair and Ju 1996, Owen \etal 1998, Ferrari \etal
1999a, 1999b, Schutz 1999, Giovannini 2000, Maggiore 2000,
Schneider \etal 2000, among others).

In the present study we have considered the background of GWs
produced from a Population III black hole formation. The basic
arguments in favour of the existence of these pre-galactic black
holes are the following: (a) from the Gunn-Peterson effect (Gunn
and Peterson 1965), it is widely accepted that the Universe
underwent a reheating (or reionization) phase between the standard
recombination epoch (at $z\sim 1000$) and $z
> 5$ as a result of the formation of the first structures of
the Universe (see, Haiman and Loeb 1997, Loeb and Barkana 2001 for
a review). However, at what redshift the reionization occured is
still an open question (Loeb and Barkana 2001), although recent
studies conclude that it occurred at redshifts in the range
$6<z<30$ (Venkatesan 2000); (b) the metallicity of $\sim 10^{-2}
Z_{\odot}$ found in ${\rm high}-z$ Ly$\alpha$ forest clouds
(Songaila and Cowie 1996, Ellison \etal 2000) is consistent with a
stellar population formed at $z > 5$ (Venkatesan 2000).

In the present paper we have adopted a stellar generation with a
Salpeter initial mass function (IMF) as well as different stellar
formation epochs. We then discuss what conclusions would be drawn
from whether (or not) the stochastic background studied here is
detected by forthcoming GW observatories such as LIGO and VIRGO.

The paper is organized as follows. In section 2 we describe how to
calculate the background of GWs produced during the formation of
the stellar black holes in this scenario (the reader finds a more
detailed description in de Araujo \etal 2002), in section 3 we
present some numerical results and the discussions, in section 4
we consider the detectability of this putative GW background and
finally in section 5 we present our conclusions.

\section{The gravitational wave production from pre-galactic stars}

Before going into detail of the calculation of the background of
GWs produced by pre-galactic stars, it is import to consider in
what ways the present study differs from previous ones.

Ferrari \etal (1999a), for example, consider the background of GWs
produced by the formation of black holes of galactic origin, i.e.,
those formed at redshifts $z<5$. Schneider \etal (2000) study the
collapse of Very Massive Objects (VMO), and the eventual formation
of very massive black holes and the GWs they generate. Fryer \etal
(2002) study, among other issues, the collapse of very massive
Population III stars ($>300{\rm M}_\odot$) and the GWs generated
by them.

Here, we consider a different approach to the formation of
pre-galactic objects. We take into account the formation of
Population III stellar black holes, considering that the
progenitor stars follow a Salpeter's IMF. In this case, the
progenitor stars have masses in the range $25-125 M_\odot$. We
also consider that the formation of these black holes occurs at
high redshifts, namely, $z>10$. It is worth noting that Sapeter's
Population III stars could account for the metallicity found in
high-z Ly$\alpha$ forest clouds, and, at least in part, since the
QSOs could also be important, for the reionization of the
Universe.

Let us now focus on how to calculate the background of GWs
produced by the Population III stellar black holes we propose
exist.

The spectral energy density, the flux of GWs, received on Earth,
$F_\nu$, in ${\rm erg}\, {\rm cm}^{-2}\,{\rm s}^{-1}\,{\rm
Hz}^{-1}$, reads (see, e.g. Douglass and Braginsky 1979, Hils
\etal 1990)

\begin{equation}
F_{\nu} = {c^{3}s_{\rm h}\omega_{\rm obs}^{2}\over 16{\rm \pi} G},
\label{fluxa}
\end{equation}

\noindent where $\omega_{\rm obs}=2{\rm \pi} \nu_{\rm{obs}}$, with
$\nu_{\rm{obs}}$ being the GW frequency (Hz) observed on Earth,
$c$ is the velocity of light, $G$ is the gravitational constant
and $\sqrt{s_{\rm h}}$ is the strain amplitude of the GW ($\rm
Hz^{-1/2}$).

The stochastic GW background produced by gravitational collapses
that lead to black holes would have a spectral density of the flux
of GWs and strain amplitude also related to equation
(\ref{fluxa}). The strain amplitude at a given frequency, at the
present time, is the contribution of black holes with different
masses at different redshifts. Thus, the ensemble of black holes
formed produces a background whose characteristic strain amplitude
at the present time is $\sqrt s_{\rm h}$.

On the other hand, the spectral density of the flux can be written
as (Ferrari \etal 1999a)

\begin{equation}
F_{\nu}=\int_{z_{\rm {cf}}}^{z_{\rm {ci}}} \int_{m_{\rm
{min}}}^{m_{\rm {u}}} f_{\nu}(\nu_{\rm{obs}}) dR_{\rm BH}(m,z)
\end{equation}

\noindent where $f_{\nu}(\nu_{\rm{obs}})$ is the energy flux per
unit of frequency (in ${\rm erg}\,{\rm cm}^{-2}\,{\rm Hz}^{-1}$)
produced due to the formation of a unique black hole and $dR_{\rm
BH}$ is the differential rate of black hole formation.

The above equation takes into account the contribution of
different masses that collapse to form black holes occurring
between redshifts $z_{\rm ci}$ and $z_{\rm cf}$ (beginning and end
of the star formation phase, respectively) that produce a signal
at the same frequency $\nu_{\rm{obs}}$. On the other hand, we can
write $f_{\nu}(\nu_{\rm{obs}})$ (Carr 1980) as

\begin{equation}
f_{\nu}(\nu_{\rm{obs}}) = {{\rm \pi} c^{3}\over 2G}h_{\rm BH}^{2}
\end{equation}

\noindent where $h_{\rm BH}$ is the dimensionless amplitude
produced by the collapse, to a black hole, of a given star with
mass $m$ that generates at the present time a signal with
frequency $\nu_{\rm{obs}}$. Then, the resulting equation for the
spectral density of the flux is

\begin{equation}
F_{\nu} = {\pi c^{3}\over 2G} \int h_{\rm  BH}^{2}dR_{\rm BH}.
\end{equation}

\par\noindent From the above equations, we obtain for the strain
amplitude

\begin{equation}
s_{\rm h} = {1 \over \nu_{\rm obs}^{2}}\int h_{\rm BH}^{2} dR_{\rm
BH}.
\end{equation}

Then, the dimensionless amplitude of the background of GWs,
$h_{\rm BG}$, reads

\begin{equation}
h_{\rm BG}^{2} = {1 \over \nu_{\rm obs}}\int h_{\rm BH}^{2}
dR_{\rm BH}
\end{equation}

\par\noindent (see, de Araujo \etal 2000 for details) where
$\nu_{\rm obs}$, $h_{\rm BH}$ and $dR_{\rm BH}$ are defined below.

The dimensionless amplitude $h_{\rm BH}$ produced by the collapse
of a star, or star cluster, to form a black hole is (Thorne 1987)

\begin{equation}
h_{\rm BH} \simeq 7.4\times 10^{-20}\, \varepsilon_{\rm
GW}^{1/2}\bigg({M_{\rm r}\over {\bf M}_{\odot}}\bigg)
\bigg({d_{\rm L}\over 1{\rm Mpc}}\bigg)^{-1} \label{hBH}
\end{equation}

\noindent where $\varepsilon_{\rm GW}$ is the efficiency of
generation of GWs, $M_{\rm r}$ is the remnant black hole mass and
$d_{\rm L}$ is the luminosity distance to the source.

It is worth mentioning that equation (\ref{hBH}) refers to the
black hole `ringing', which has to do with the de-excitation of
the black hole quasi-normal modes. Note also that
$\varepsilon_{\rm GW} \propto a^{4}$ (see, e.g. Stark and Piran
1986), where `$a$' is the the dimensionless angular momentum.
Thus, greater the GW efficiency, greater the dimensionless angular
momentum.

We assume that the progenitor masses of the black holes range from
$M=25-125{\rm M}_\odot$ (see Timmes \etal 1995, Woosley and Timmes
1996). The remnant and the progenitor masses are related to
$M_{\rm r}=\alpha M$ and, we assume $\alpha=0.1$ (see, e.g.
Ferrari \etal 1999a).

The collapse of a star to a black hole produces a signal with an
observed  frequency $\nu_{\rm{obs}}$ at the Earth (Thorne 1987)

\begin{equation}
\nu_{\rm{obs}} \simeq 1.3\times 10^{4}\,{\rm Hz}\bigg({{\rm
M}_{\odot}\over M_{\rm r}}\bigg)(1+z)^{-1}\label{freq}
\end{equation}

\noindent where the factor $(1+z)^{-1}$ takes into account the
redshift effect on the emission frequency.

The differential rate of black hole formation $dR_{\rm BH}$ reads

\begin{equation}
dR_{\rm BH} = \dot\rho_{\star}(z) {dV\over dz} \phi(m)dmdz
\end{equation}

\noindent where $dV$ is the co-moving volume element, $\phi(m)$ is
the stellar initial mass function (IMF) and $\dot\rho_{\star}(z)$
is the star formation rate (SFR) density.

The SFR density  can be related to the reionization of the
Universe. The amount of baryons necessary to participate in early
star formation, to account for the reionization, would amount to a
small fraction, $f_{\star}$, of all baryons of the Universe (see,
e.g. Loeb and Barkana 2001). We then assume that

\begin{equation}
\dot\rho_\star \equiv {d\rho_{\star}\over dt}= {d\over dt}
[\Omega_{\star}\;\rho_{\rm c}\;(1+z)^{3}]
\end{equation}

\par\noindent where the term in brackets
represents the stellar mass density at redshift $z$, with
$\rho_{\rm c}$ the present critical density, and $\Omega_{\star}$
the stellar density parameter. The latter can be written as a
fraction of the baryonic density parameter, namely,
$\Omega_{\star}=f_{\star}\Omega_{\rm B}$, which we assume to be
independent of the redshift.

From the above equations, we obtain for the dimensionless
amplitude

\begin{eqnarray}
h_{\rm BG}^{2} &=& {(7.4\times 10^{-20}\alpha)^{2}\varepsilon_{\rm
GW} \over \nu_{\rm{obs}}} \nonumber\\ && \times \bigg[\int_{z_{\rm
cf}}^{z_{\rm ci}} \int_{m_{\rm min}}^{m_{\rm u}}\bigg({m\over {\rm
M}_{\odot}}\bigg)^{2}\bigg({d_{\rm L}\over 1{\rm Mpc}}\bigg)^{-2}
   \dot\rho_{\star}(z){dV\over
dz} \phi(m)dmdz\bigg]\label{hBG}
\end{eqnarray}

\noindent where $m_{\rm min}=25{\rm M}_\odot$,  $m_{\rm u}=125{\rm
M}_\odot$ and $z_{\rm ci}$ ($z_{\rm cf}$) is the beginning (end)
of the black hole formation phase. Equation (\ref{hBG}) is
computed for each observed frequency. Also, looking at equation
(\ref{hBG}), one notes that to integrate it, one needs to choose
the IMF, the cosmological parameters and set values for the
following parameters: $z_{\rm ci}$, $z_{\rm cf}$, $\alpha$,
$\varepsilon_{\rm GW}$, $f_{\star}$. In the next section we
present the numerical results and discussions.

\section{Numerical results and discussions}

To evaluate the background of GWs produced by the formation of the
Population III black holes, it is necessary to know in which
redshifts they began and finished formation. This is a very
difficult question to answer, since it involves knowledge of the
role of the negative and positive feedbacks of star formation
which are regulated by cooling and injection of energy processes.

Should the stochastic background of GWs studied here be
significantly produced and detected at a reasonable confidence
level, the present study can be used to obtain the redshift range
where the Population III black holes were formed. We refer the
reader to the paper by de Araujo \etal (2002) for further
discussions. In figure 1 an example is given of how one could get
$z_{\rm ci}$ and $z_{\rm cf}$ from the curve $h_{\rm BG}$ versus
$\nu_{\rm obs}$. Knowing the frequency band $\nu_{\rm
min}-\nu_{\rm max}$ detected from a cosmological source, and using
equation (\ref{freq}), one can obtain both $z_{\rm ci}$ and
$z_{\rm cf}$ from figure 1. Thus, these redshifts are therefore
observable. Note that we have assumed as did Ferrari \etal (1999a)
that $\alpha$ is a constant ($\alpha =0.1$).

\begin{figure}
\begin{center}
\leavevmode
\centerline{\epsfig{figure=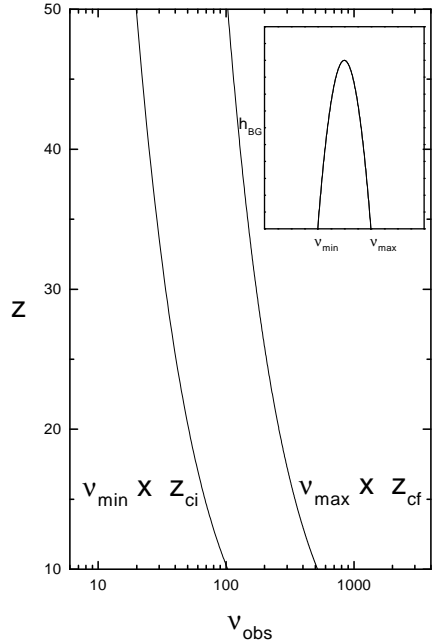,angle=360,height=10cm,width=7cm}}
\caption{Example of how one could obtain the initial (final)
redshift  $z_{\rm ci}$ ($z_{\rm cf}$) of the GW emission period
from $h_{\rm BG}$ versus $\nu_{\rm obs}$. We have adopted
$\alpha=0.1$.}
\end{center}
\end{figure}

Stars start forming at different redshifts, creating ionized
bubbles (Str\"{o}mgren spheres) around themselves, which expand
into the intergalactic medium (IGM), at a rate dictated by the
source luminosity and the background IGM density (Loeb and Barkana
2001). The reionization is complete when the bubbles overlap to
fill the entire Universe. Thus the epoch of reionization is not
the epoch of star formation. There is a non-negligible time span
between them. Here, we have chosen different formation epochs to
see their influence on the putative background of GWs and also to
see if it could be detected by the forthcoming GW antennas.

To calculate $h_{\rm BG}$ we adopted the standard Salpeter IMF.
For $\varepsilon_{\rm GW}$, the efficiency of production of GWs,
whose distribution function is unknown, we have parametrized our
results in terms of its maximum value, namely, $\varepsilon_{\rm
GW_{\rm max}}=7 \times 10^{-4}$. This figure is obtained from
studies by Stark and Piran (1986) who simulated the axisymmetric
collapse of a rotating star to a black hole.

To calculate $h_{\rm BG}$ we still need to know $\Omega_{\star}$,
which has a key role in the definition of the SFR density. From
different studies one can conclude that a few per cent, may be up
to $\sim 10\%$, of the baryons must be condensed into stars in
order for the reionization of the Universe to take place (see,
e.g. Venkatesan 2000). Here we have set the value of
$\Omega_{\star}$ in such a way that it amounts to $1\%$ of all
baryons (our fiducial value).

Looking at equation (\ref{hBG}), one could think it would depend
critically on the cosmological parameters $H_0$, $\Omega_{\rm B}$,
$\Omega_{\rm DM}$ (the density parameter for the dark matter) and
$\Omega_\Lambda$ (the density parameter associated with the
cosmological constant). But our results show that $h_{\rm BG}$
depends only on $H_{0}$ and $\Omega_{\rm B}$, the Hubble parameter
and the baryonic density parameter, respectively.

The quantity $h_{100}^{2}\Omega_{\rm B}=0.019\pm 0.0024$ (where
$h_{100}$ is the Hubble parameter given in terms of 100 ${\rm
km\;s^{-1}\;Mpc^{-1}}$) is adopted in the present model. This
figure is obtained from Big Bang nucleosynthesis studies (see,
e.g. Burles \etal 1999).

In table 1 we present the redshift band, $z_{\rm ci}$ and $z_{\rm
cf}$, for the models studied and the corresponding GW frequency
bands. For the cosmological parameters, we have adopted
$h_{100}=0.65$, $\Omega_{\rm M}=0.3$, $\Omega_{\rm B}=0.045$ and
$\Omega_{\Lambda}=0.7$. We have also adopted $\alpha=0.1$,
$f_{\star}=0.01$ and the standard IMF.

It is worth noting that no structure formation model has been used
to find the black holes formation epoch; instead, we have simply
chosen the values of $z$ to see whether it is possible to obtain
detectable GW signals. In the next section, it will be seen that
unless $\varepsilon_{\rm GW}$ is negligible, the  background of
GWs we propose here can be detected. Our choices, however, can be
understood as follows. The greater the redshift formation, the
more power the masses related to the Population III objects have.
Thus from our models A to D, our model D (A) has more (less) power
when compared to the others. The models E, F and G would mean a
more extended star formation epoch, which means that the feedback
processes of star formation are such that they allow a more
extended star formation epoch when compared to the models B, C and
D, respectively.

Note that the reionization epoch occurred at lower redshifts as
compared to the first stars formation redshifts. Loeb and Barkana
(2001) found, for example, that if the stars were formed at $z
\simeq 10-30$, with standard IMF, they could have reionized the
Universe at redshift $z \sim 6$. Our models A, B and E, for
example, could account for such a reionization redshift.

If the process of structure formation of the Universe and the
consequent star formation were well known, one could obtain the
redshift formation epoch of the first stars. On the other hand, if
the background of GWs really exists and is detected, one can
obtain information about the formation epoch of the first stars.

\begin{table}
\caption{The redshifts of collapse for our models and the
corresponding GW frequency bands. The cosmological parameter
$h_{100}^{2}\Omega_{\rm B}=0.019$ (see the text), $\alpha=0.1$,
$f_{\star}=0.01$ (our fiducial value) and the standard IMF are
adopted.}
\begin{indented}
\item[]
\begin{tabular}{ccccccc}
\br
Model & $z_{\rm ci}$ & $z_{\rm cf}$ &  $ \Delta\nu_{\rm obs} $ (Hz) \\
\mr
A  &  20  &  10 & 50-470 \\
B  &  30  &  20 & 34-250 \\
C  &  40  &  30 & 25-170 \\
D  &  50  &  40 & 20-130 \\
E  &  30  &  10 & 34-470 \\
F  &  40  &  10 & 25-470 \\
G  &  50  &  10 & 20-470 \\
\br
\end{tabular}
\end{indented}
\end{table}

A relevant question is whether the background we study here is
continuous or not. The duty cycle indicates if the collective
effect of the bursts of GWs generated during the collapse of a
progenitor star generates a continuous background. For all the
models studied here the duty cycle is $\gg 1$ (see de Araujo \etal
2002 for details).

We find, for example, that the formation of a Population (III) of
black holes, in the model D, could generate a stochastic
background of GWs with amplitude $h_{\rm BG} \simeq (0.8-2)\times
10^{-24}$ and a corresponding closure density of
$\Omega_{\rm{GW}}\simeq (0.7-1.4)\times 10^{-8}$, at the frequency
band $\nu_{\rm{obs}} \simeq 20-130\, {\rm Hz}$ (assuming an
efficiency of generation $\varepsilon_{\rm GW} \simeq 7\times
10^{-4}$, the maximum one).

In other paper to appear elsewhere (de Araujo \etal 2002), we
study in detail how the variations of the several parameters
modify our results.

\section{Detectability of the background of gravitational waves}

The background predicted in the present study cannot be detected
by single forthcoming interferometric detectors, such as VIRGO and
LIGO (even by advanced ones). However, it is possible to correlate
the signal of two or more detectors to detect the background that
we propose to exist.

To assess the detectability of a GW signal, one must evaluate the
signal-to-noise ratio (SNR), which for a pair of interferometers
is given by (see, e.g. Flanagan 1993)

\begin{equation}
{\rm SNR}^2=\left[\left(\frac{9 H_0^4}{50\pi^4} \right) T
\int_0^\infty d\nu \frac{\gamma^2(\nu)\Omega^2_{GW}(\nu) } {\nu^6
S_{\rm h}^{(1)}(\nu) S_{\rm h}^{(2)}(\nu)} \right]
\end{equation}

\noindent where $ S_{\rm h}^{(i)}$ is the spectral noise density,
$T$ is the integration time and $\gamma(\nu)$ is the overlap
reduction function, which depends on the relative positions and
orientations of the two interferometers. The closure energy
density is given by

\begin{equation}
\Omega_{\rm GW} = {1\over \rho_{\rm c}} {d\rho_{\rm GW}\over d\log
\nu_{\rm{obs}}}={4{\rm \pi}^{2}\over 3H^{2}_{0}}\nu_{\rm{obs}}^{2}
h_{\rm BG}^{2}.
\end{equation}

Here we consider, in particular, the LIGO interferometers, and
their spectral noise densities have been taken from a paper by
Owen \etal (1998).

In table 2 we present the SNR for 1 year of observation with
$\alpha=0.1$, $\Omega_{\rm B}h^{2}_{100}=0.019$, $f_{\star}=0.01$
and $\varepsilon_{\rm GW_{\rm max}}=7\times 10^{-4}$ for the
models in table 1, for the three LIGO interferometer
configurations.

\begin{table}
\caption {For the models in table 1, we present the SNR for pairs
of LIGO I, II and III (`first', `enhanced' and `advanced',
respectively) observatories for 1 year of observation. Note that
an efficiency of generation $\varepsilon_{\rm GW_{\rm max}} =
7\times 10^{-4}$ is assumed.}
\begin{indented}
\item[]
\begin{tabular}{cccc}
\br
      &        &  SNR    &          \\ \ns \ns
      & \crule{3} \\
Model & LIGO I  & LIGO II  & LIGO III \\
\mr
A  & $8.3\times 10^{-3}$  & $1.6$  & $6.6$ \\
B  & $8.5\times 10^{-3}$  & $2.3$  & $26 $ \\
C  & $8.7\times 10^{-3}$  & $2.7$  & $47 $ \\
D  & $8.1\times 10^{-3}$  & $2.5$  & $51 $ \\
E  & $2.7\times 10^{-3}$  & $5.7$  & $37 $ \\
F  & $5.0\times 10^{-3}$  & $12 $  & $120$ \\
G  & $7.7\times 10^{-2}$  & $21 $  & $260$ \\
\br
\end{tabular}
\end{indented}
\end{table}

Note that for the `initial'  LIGO (LIGO I), there is no hope of
detecting the  background of GWs we propose here. For the
`enhanced' LIGO (LIGO II) there is some possibility of detecting
the background, since ${\rm SNR}
> 1$, if $\varepsilon_{\rm GW}$ is around the maximum value. Even
if the LIGO II interferometers cannot detect such a background, it
will be possible to constrain the efficiency of GW production.

The prospect of detection with the `advanced' LIGO (LIGO III)
interferometers is much more optimistic, since the SNR for almost
all models is significantly greater than unity. Only if the value
of $\varepsilon_{\rm GW}$ were significantly lower than the
maximum value would the detection not be possible. In fact, the
signal-to-noise ratio is critically dependent on this parameter.

Note that the larger the star formation redshift band, the greater
the SNR. Secondly, the earlier the star formation, the greater the
SNR. It is worth recalling that, if one can obtain the curve
`$h_{\rm BG}$ versus $\nu_{\rm obs}$' and the value of $\alpha$ is
known, one can find the redshift of star formation.

\section{Conclusions}

We have shown that a background of GWs is produced from Population
III black hole formation at high redshift. This background can in
principle be detected by a pair of LIGO II (or more probably by a
pair of LIGO III) interferometers. However, a relevant question
should be considered: what astrophysical information can one
obtain whether or not such a putative background is detected?

First, let us consider a non-detection of the GW background. The
critical parameter to be constrained here is $\varepsilon_{\rm
GW}$. A non-detection would mean that the efficiency of GWs during
the formation of black holes is not high enough. Another
possibility is that the first generation of stars is such that the
black holes formed had masses $> 100M_{\odot}$, and should they
form at $z > 10$ the GW frequency band would be out of the LIGO
frequency band.

Secondly, a detection of the background with a significant SNR
would permit us to obtain the curve of $h_{\rm BG}$ versus
$\nu_{\rm obs}$. From it, one can constrain $\alpha$ and the
redshift formation epoch; and for a given IMF and $\Omega_{\rm
B}h^{2}$, one can also constrain the values of $f_{\star}$ and
$\varepsilon_{\rm GW}$. On the other hand, using the curve of
$h_{\rm BG}$ versus $\nu_{\rm obs}$ and in addition other
astrophysical data, say CBR data, models of structure formation
and reionization of the Universe, constraint on $\varepsilon_{\rm
GW}$ can also be imposed. We refer the reader to the paper by de
Araujo \etal (2002) for further discussions.

It is worth mentioning that a significant amount of GWs can also
be produced during the formation of neutron stars and if such
stars are r-mode unstable (Andersson 1998).  We leave these issues
for other studies to appear elsewhere.

\ack The authors thank FAPESP (Brazil) for support (grant numbers
97/06024-4, 97/13720-7, 98/13468-9, 98/13735-7, 00/00116-9,
01/04086-0, and 01/04189-3). JCNA and ODA also thank CNPq (Brazil)
for support (grant numbers 450994/01-5 and 300619/92-8,
respectively). We also thank the referees for the suggestions and
criticisms.

\References

\item[]Andersson N 1998 { \it Astrophys. J.} {\bf
211} 708

\item[]Blair D G and Ju L 1996 {\it Mon. Not. R. Astron. Soc.} {\bf
283} 618

\item[]Burles S, Nollett K M, Truran J N
and Turner M S 1999 {\it Phys.Rev.Lett.} {\bf 82} 4176

\item[]Carr B J 1980 {\it Astron. Astrophys.} {\bf 89} 6

\item[]de Araujo J C N, Miranda O D and Aguiar O D 2000 {\it Phys. Rev.} D
{\bf 61} 124015

\item[]de Araujo J C N, Miranda O D and Aguiar O D 2002 {\it Mon. Not. R. Astron. Soc.}
at press

\item[]Douglass D H and Braginsky  V G 1979 {\it General
Relativity: An Einstein Centenary Survey} (Cambridge: Cambridge
University Press) p~90

\item[]Ellison S, Songaila A, Schaye J and Petinni M 2000 {\it Astron. J.} {\bf 120} 1175

\item[]Ferrari V, Matarrese S and Schneider R 1999a {\it Mon. Not. R. Astron. Soc.} {\bf 303} 247

\item[]Ferrari V, Matarrese S and Schneider R 1999b {\it Mon. Not. R. Astron. Soc.} {\bf 303} 258

\item[]Flanagan E E 1993 {\it Phys. Rev.} D {\bf 48}  2389

\item[] Fryer C L, Holz D E and Hughes S A 2002 { \it Astrophys. J.} {\bf
565} 430

\item[]Giovannini M 2000 {\it Preprint} gr-qc/0009101

\item[] Gunn J E and Peterson B A 1965 { \it Astrophys. J.} {\bf 142}
1633

\item[]Haiman Z and Loeb A 1997 {\it Astrophys. J.} {\bf 483} 21

\item[]Hils D, Bender P L and Webbink R F 1990 { \it Astrophys.
J.} {\bf 360} 75

\item[]Loeb A and Barkana R 2001 {\it Ann. Rev. Astron. Astrophys.} {\bf
39} 19

\item[]Maggiore M 2000 {\it Phys. Rep.} {\bf 331} 283

\item[]Owen B J, Lindblom L, Cutler C, Schutz B F, Vecchio A and Andersson N 1998
{\it Phys.  Rev.} D {\bf 58} 084020

\item[]Schneider R, Ferrara A, Ciardi B, Ferrari V and Matarrese S 2000 {\it Mon. Not. R.
Astron. Soc.} {\bf 317} 385

\item[]Schutz B F 1999 {\it Class. Quantum Grav.} {\bf 16} A131

\item[]Songaila A and Cowie L L 1996 {\it Astron. J.} {\bf 112} 335

\item[]Stark R F and Piran T 1986 {\it Proc. 4th Marcel Grossmann
Meeting on General Relativity (Rome, Italy)} (Amsterdam: Elsevier)
p 327

\item[]Thorne K S 1987 {\it 300 years of Gravitation} (Cambridge: Cambridge
University Press) p~331

\item[] Timmes F X, Woosley S E and Weaver T A 1995 {\it Astrophys. J. Suppl } {\bf 98}
617

\item[]Venkatesan A 2000 {\it Astrophys. J.} {\bf 537} 55

\item Woosley S E and Timmes F X 1996 {\it Nucl. Phys.} A {\bf
606} 137

\endrefs
\end{document}